\documentclass[aps,prl,superscriptaddress,twocolumn,showpacs,floatfix]{revtex4-1}
\usepackage{graphicx}
\usepackage{subfigure}
\usepackage{color}
\usepackage{dcolumn}
\usepackage{bm}
\newcommand{\Boc}{Bo$_{\hbox{\rm c}}$~}
\newcommand{\Bocm}{Bo$_{\hbox{\rm c}}$}

\begin{document}
\title{Field-Induced Breakup of Emulsion Droplets Stabilized by Colloidal Particles}

\author{E. Grace Kim}
 \affiliation{Max-Planck-Institute for Intelligent systems, Heisenbergstr. 3, 70569 Stuttgart, Germany}
\author{Kevin Stratford}
\affiliation{Edinburgh Parallel Computing Centre, The University of Edinburgh, JCMB, The King's Buildings, Mayfield Road, Edinburgh EH9 3JZ, UK}
\author{Paul S. Clegg} \author{Michael E. Cates}
\affiliation{SUPA, School of Physics and Astronomy, The University of Edinburgh, JCMB, The King's Buildings, Mayfield Road, Edinburgh EH9 3JZ, UK}

 \date{\today}
 \begin{abstract}
 We simulate the response of a particle-stabilized emulsion droplet in an external force field, such as gravity, acting equally on all $N$ particles.  We show that the field strength required for breakup (at fixed initial area fraction) decreases markedly with droplet size, because the forces act cumulatively, not individually, to detach the interfacial particles. The breakup mode involves the collective destabilization of a solidified particle raft occupying the lower part of the droplet, leading to a critical force per particle that scales approximately as $N^{-1/2}$.
 \end{abstract}
 \maketitle


The stabilization of emulsions is an important issue in industry, where it controls the functionality and shelf-life of many products \cite{miller:2008/a}. Alongside traditional surfactants, colloidal particles or nanoparticles have long been used to stabilize emulsion droplets (Pickering emulsions) \cite{pickering:1907/a}, with several emerging applications that exploit the nonequilibrium nature of their adsorption to the fluid-fluid interface \cite{Leal,binksnat,lin}. However this also poses obstacles to formulation: to maintain emulsion stability, it is essential that the colloidal particles remain sequestered on the interface, even when the system is subject to external challenges \cite{Aveyard,Leal}. 

Brownian motion is generally insufficient to cause detachment \cite{Aveyard,Kevin}, so we ignore it here. In contrast, external body forces acting equally on each particle can, if strong enough, overcome the capillary forces holding them in place. Destabilizing forces can be caused by magnetic or electric fields or their gradients, and also by gravity. For simplicity we refer mainly to gravity in what follows.

A groundbreaking experimental study was performed by Melle {\it et al.}~\cite{fuller:2005/a}, using decane droplets in water stabilized by paramagnetic particles. They demonstrated, using a simple bar magnet, controlled field-gradient induced breakup of an otherwise highly stable emulsion.  A second study uses instead the dielectrophoretic force caused by an electric field gradient \cite{aubry:2010/a}. A third recent study has found complex dynamics in droplets stabilized by supra-colloidal (400$\mu$m) particles under gravity \cite{Joe}. (Scaling by the Bond number, defined below, indicates that the same physics should arise for, e.g., 1$\mu$m particles under bench-top centrifugation.) Although field-induced breakup is simultaneously a threat to an emulsion's stability and a promising route to its controlled breakage, the dependence of the critical field for breakup on the droplet size is far from understood.

Previously we performed lattice Boltzmann (LB) simulations on emulsion structures (primarily bicontinuous \cite{Kevin,Paul}) stabilized by magnetic colloids in the presence of a magnetic field gradient \cite{kim:one,kim:two}. There we argued that the
effect of the force does not act separately on each particle but is cumulative across the structure. (This explains the very low field-gradients required in \cite{fuller:2005/a}.) By that argument, a major fraction of the weight of all particles 
could be loaded onto a single `keystone' particle at the droplet bottom which should then detach, initiating breakup. 
In this paper, we study in detail the breakup dynamics of isolated emulsion droplets coated with $N$ (essentially hard-sphere) particles under gravity. 
We find a scaling of the critical force with $N$ that is inconsistent with the `keystone' argument, and associated instead with a failure mode in which a raft of jammed particles detaches from the droplet base taking liquid with it.

\textit{Control parameters:}
For a single particle adsorbed onto a flat horizontal interface between two fluids of equal density, detachment is controlled by the Bond number \cite{clift:1979/a}
${\rm Bo}= {3}{F}/ {(4\pi\sigma a)} $
where $a$ is the particle radius, $\sigma$ the fluid-fluid interfacial tension, and $F$ the normal detachment force (in gravity, $F$ is the buoyancy force on the particle).
Detachment of an isolated particle ($N=1$) occurs at a critical Bond number \Bocm $(1)$ which depends on the fluid-solid interfacial tensions through the contact angle $\theta$. In this work we consider only neutral wetting ($\theta  = 90^\circ$) for which \Bocm $(1)= 3/4$ \cite{stratford:2005/a,derjaguin:1946/a}. The critical Bond number for detachment of one particle from a fluid sphere of radius $R$ should also depend on $r\equiv R/a$; to allow for this we define \Bocm $(1,r)$ such that \Bocm $(1,\infty) =$ \Bocm $(1)$.  The numerically determined \Bocm $(1,r)$ additionally reflects various sources of discretization error (e.g., the fact that in LB the particle radius is not much larger than the interfacial width \cite{supmat}).

\textit{Simulations:}
The LB methodology used in this work is by now standard \cite{kim:one,stratford:2005/a,Bounce} and summarized in \cite{supmat}.
The initial condition in our simulations comprises a fluid droplet of radius $R$, immersed in another fluid of equal mass density, with $N$ neutrally wetting particles present on its surface. We choose a near-constant surface area fraction $\alpha=N\pi a^2/4\pi R^2 = N/ 4r^2 = 0.53-0.54$; this is high enough to represent a stable Pickering emulsion droplet but low enough to allow particles to rearrange on application of even a weak body force. The chosen $N$ values are $N=42,50,102,162,194$ and $700$.
Each droplet is placed in a box, with periodic boundary conditions (a repeat with closed-box boundaries for $N=700$ gave near-identical results), whose LB lattice size (depending on $R$) lies between $64^2 \times 128$ and $200^2\times 400$. 

We have examined three different initial arrangements of particles: one disordered, one a regular hexagonal packing and one a regular square packing (subject to the defects required to tile these onto a sphere). For the ordered configurations, initial particle coordinates lie on the vertices of a suitable polyhedron. 
The particles and droplet are then relaxed in the absence of body forces ($F=0$) to allow the interfacial structure to equilibrate.
Once this is done, a vertical body force $F$ is switched on; this acts directly on each particle. To prevent overall translation of the droplet's center of mass, a balancing buoyancy force $NF$ is applied as a uniform upward force density
on the fluid nodes occupied by the droplet interior. The structure of the droplet is then evolved by the LB algorithm which faithfully includes the effects of fluid flow and hydrodynamic interactions.

\begin{figure}[tbph]
\centering
\subfigure[]{\label{fig:order_initial}\includegraphics[height=0.2\textwidth]
{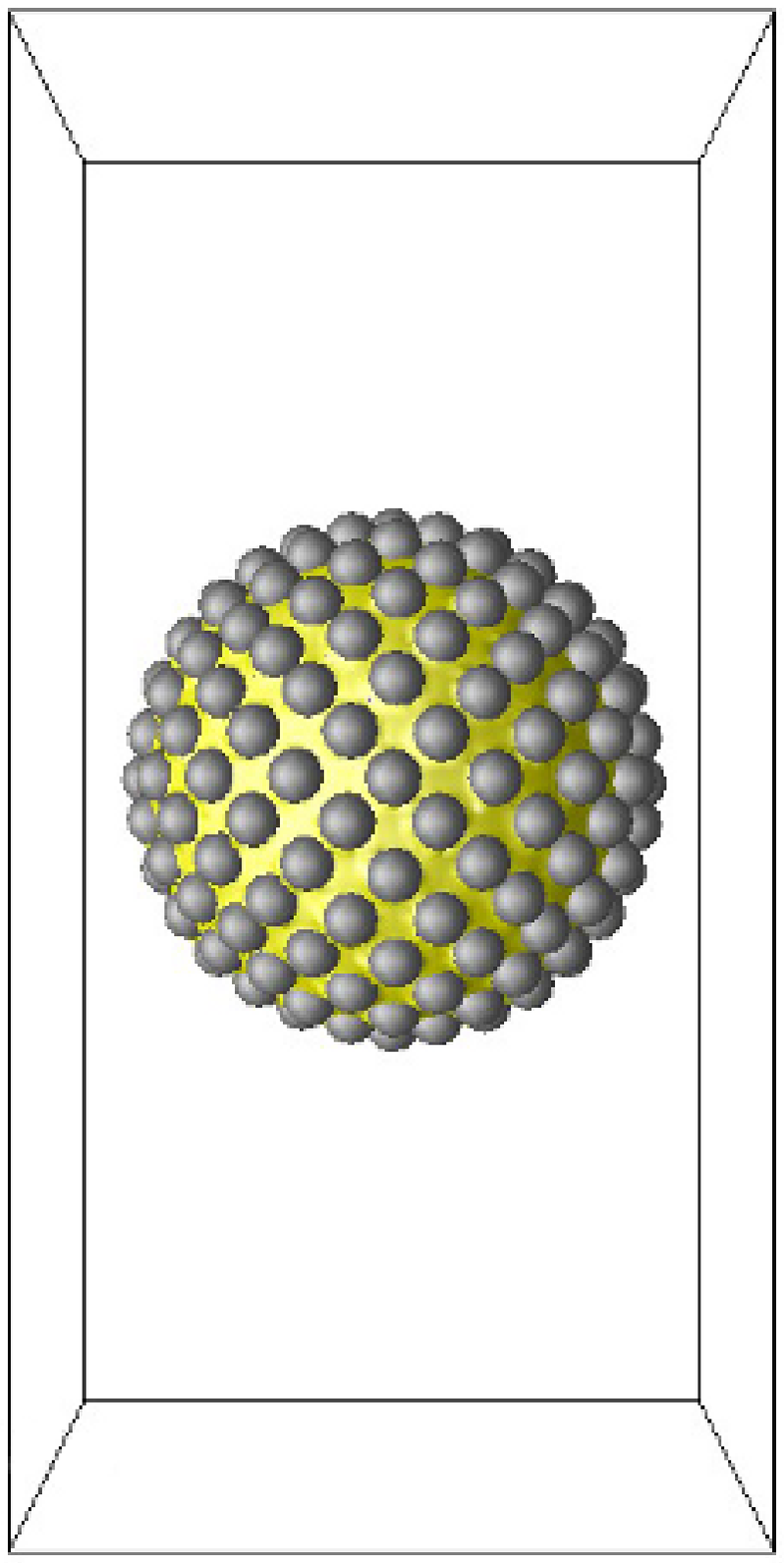}}
\subfigure[]{\label{fig:order_stable01}\includegraphics[height=0.2\textwidth]
{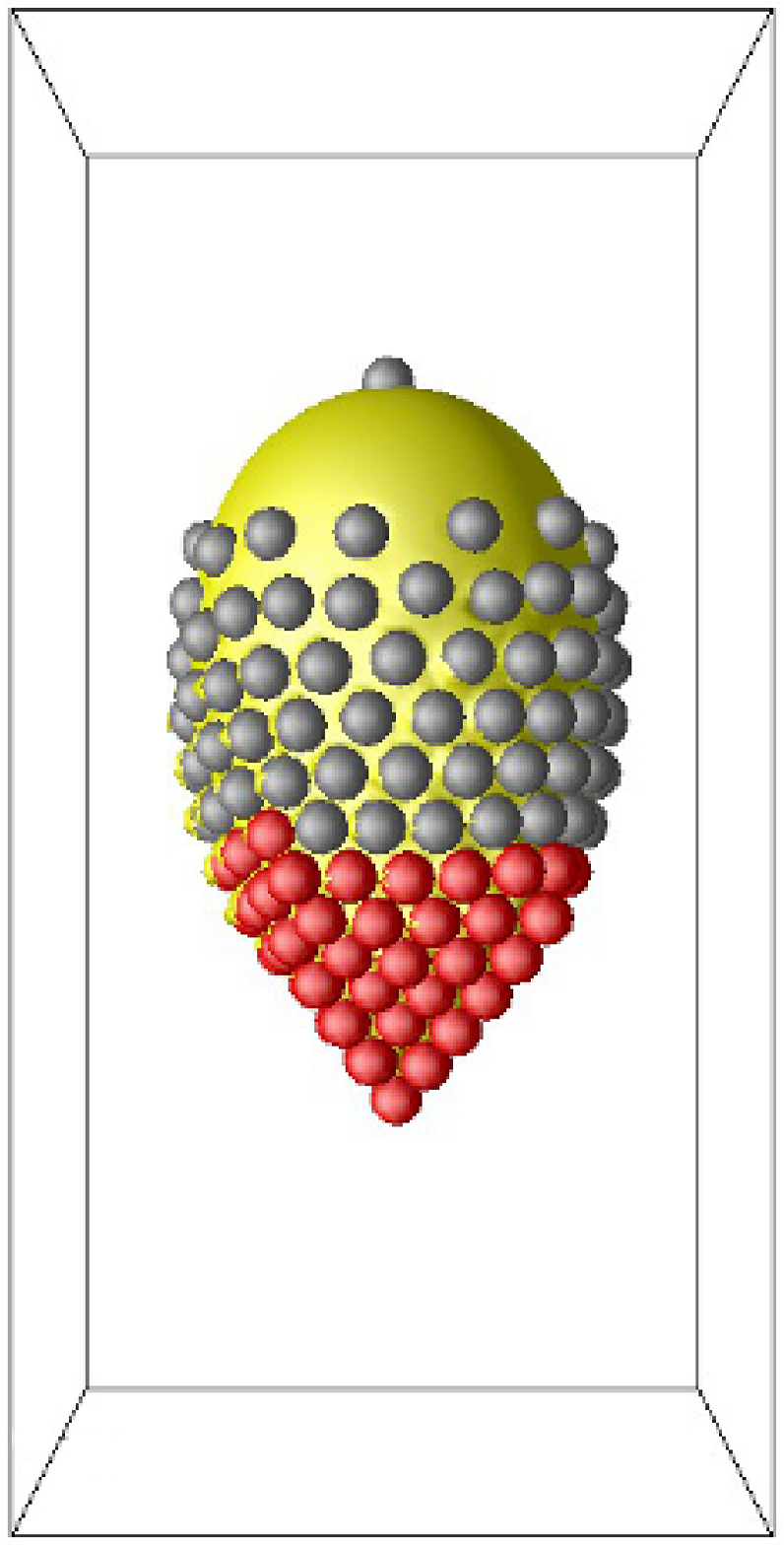}}
\subfigure[]{\label{fig:order_stable02}\includegraphics[height=0.2\textwidth]
{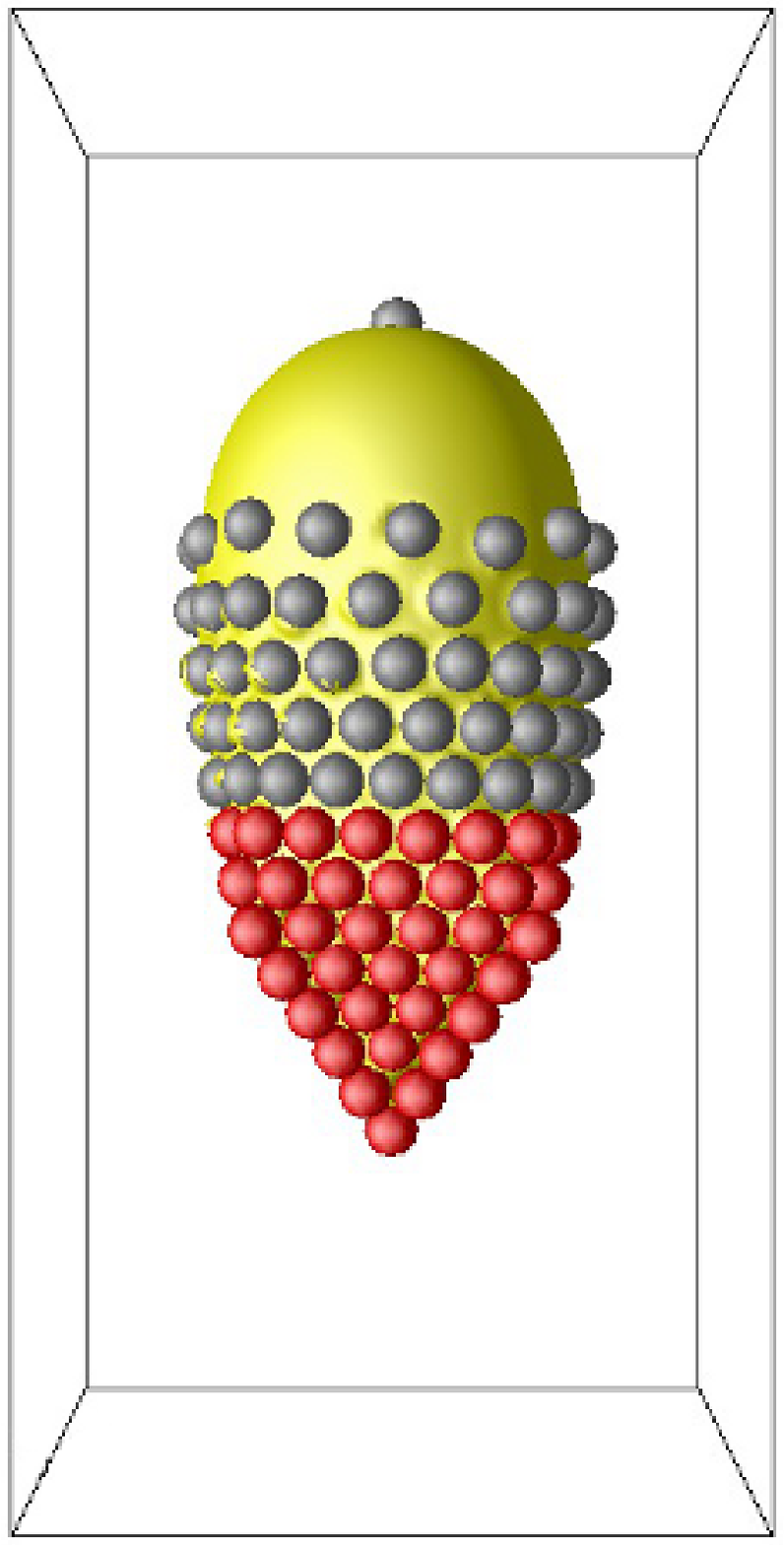}}
\subfigure[]{\label{fig:order_unstable}\includegraphics[height=0.2\textwidth]
{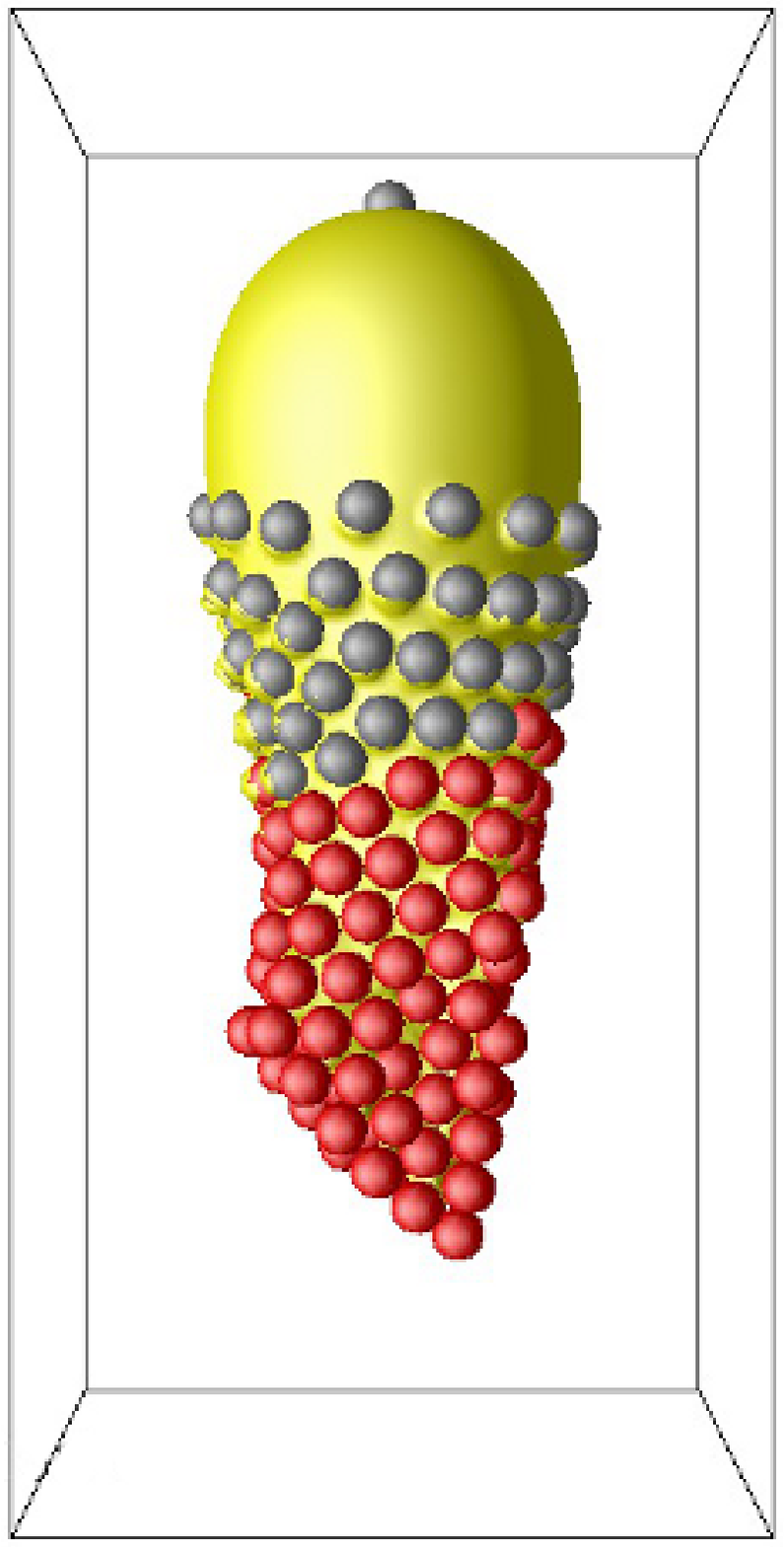}}\\
\subfigure[]{\label{fig:disorder_initial}\includegraphics[height=0.2\textwidth]
{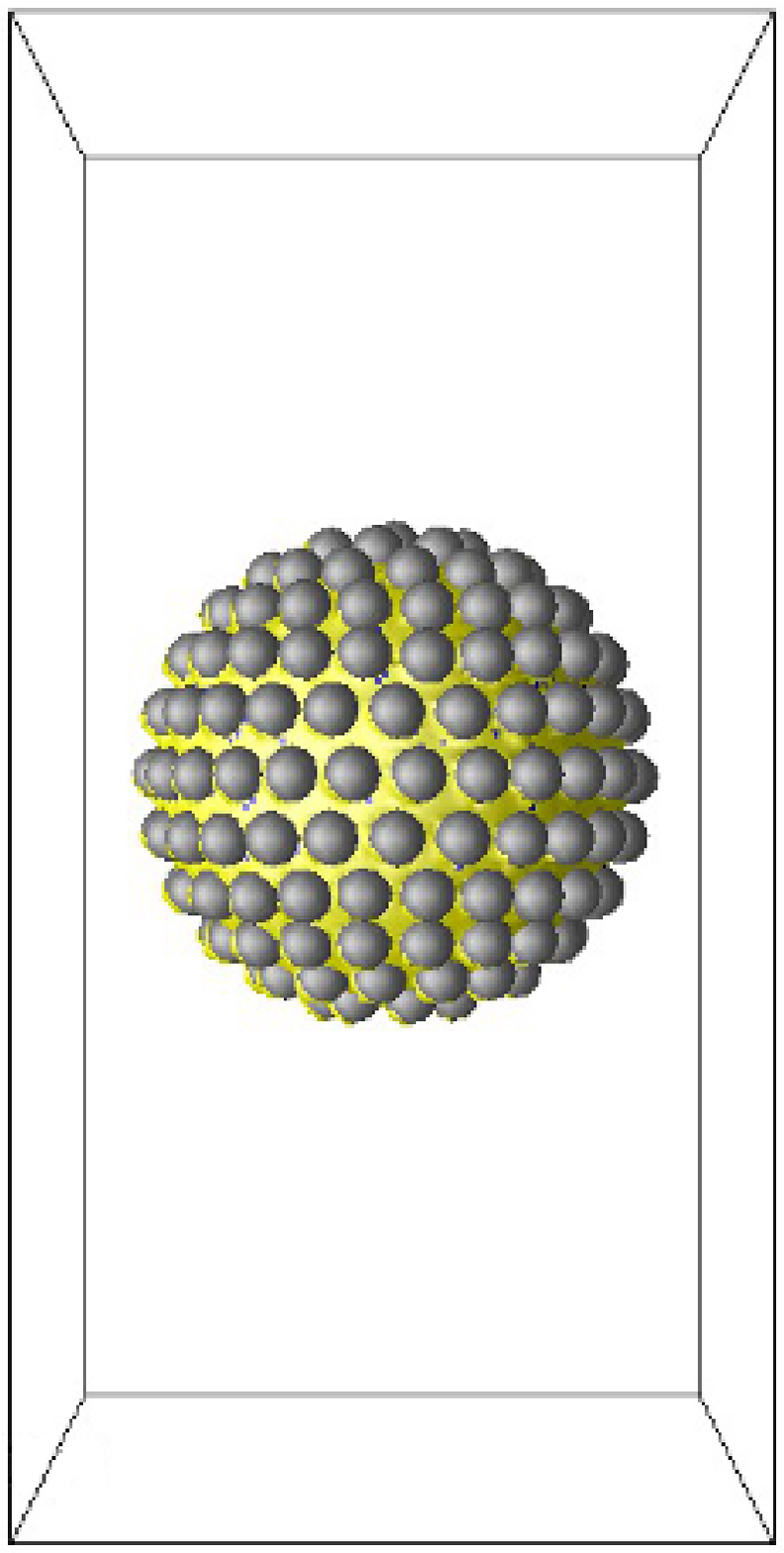}}
\subfigure[]{\label{fig:disorder_stable01}
\includegraphics[height=0.2\textwidth]{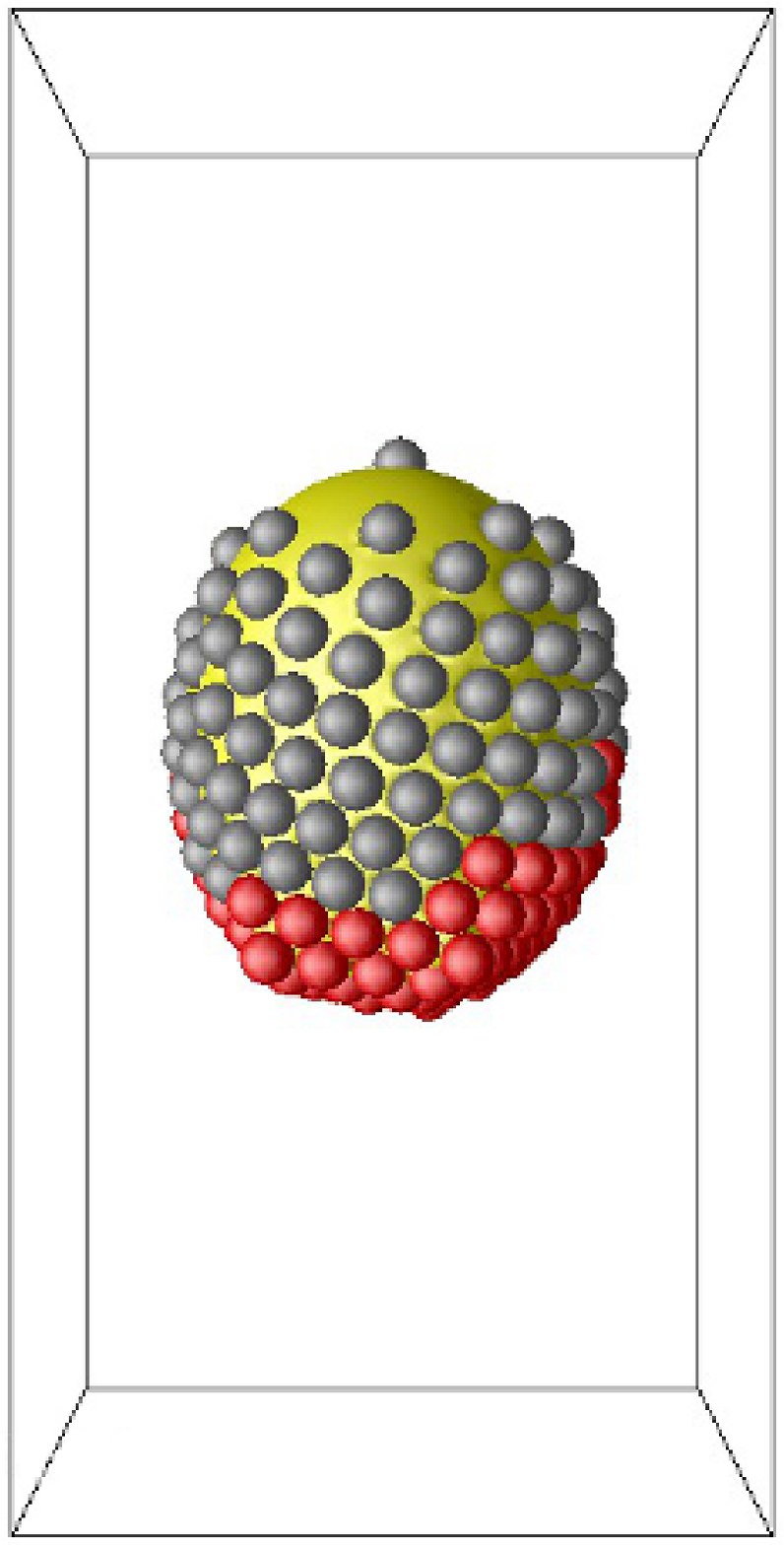}}
\subfigure[]{\label{fig:disorder_stable02}
\includegraphics[height=0.2\textwidth]{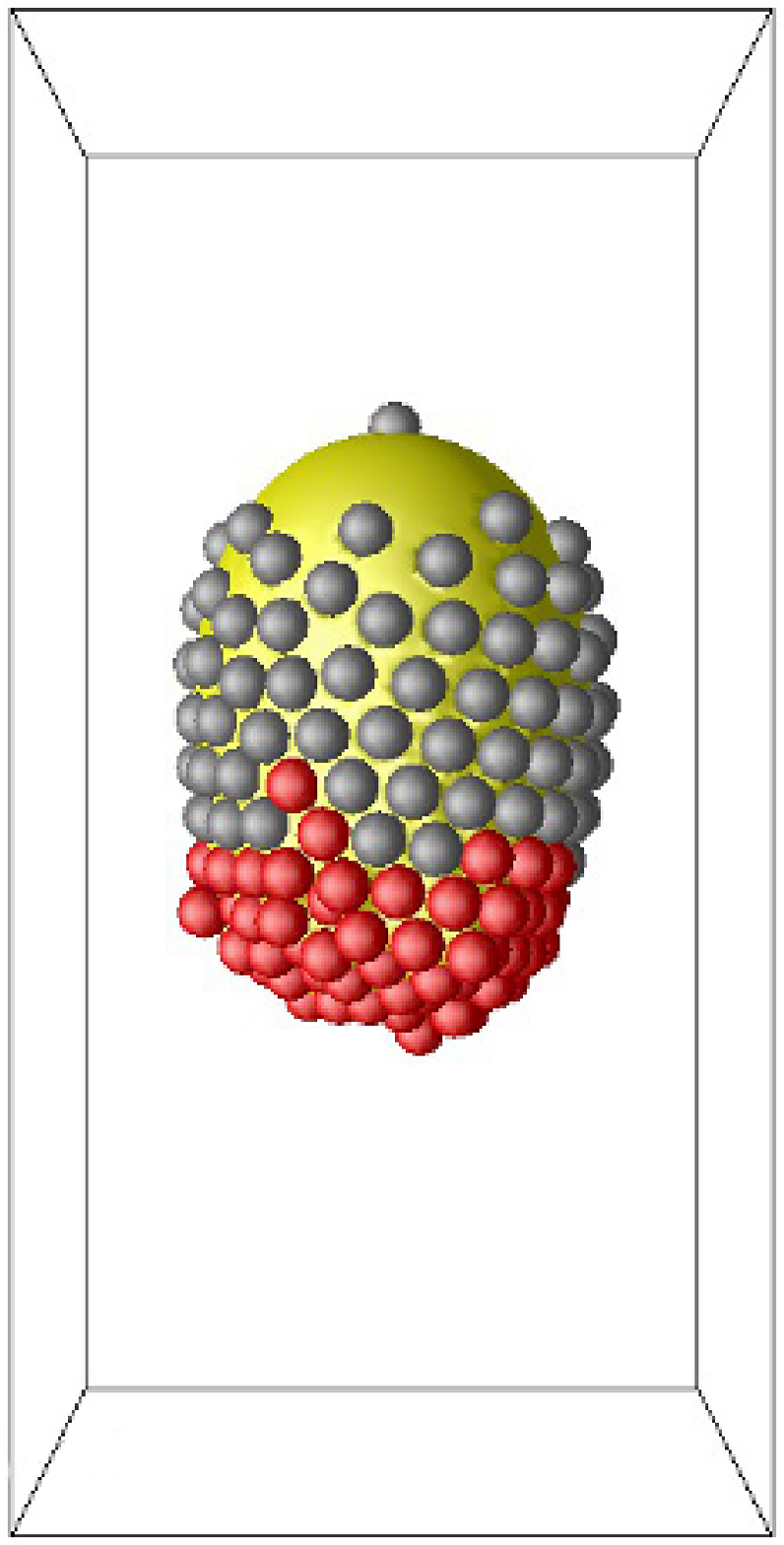}}
\subfigure[]{\label{fig:disorder_unstable}
\includegraphics[height=0.2\textwidth]{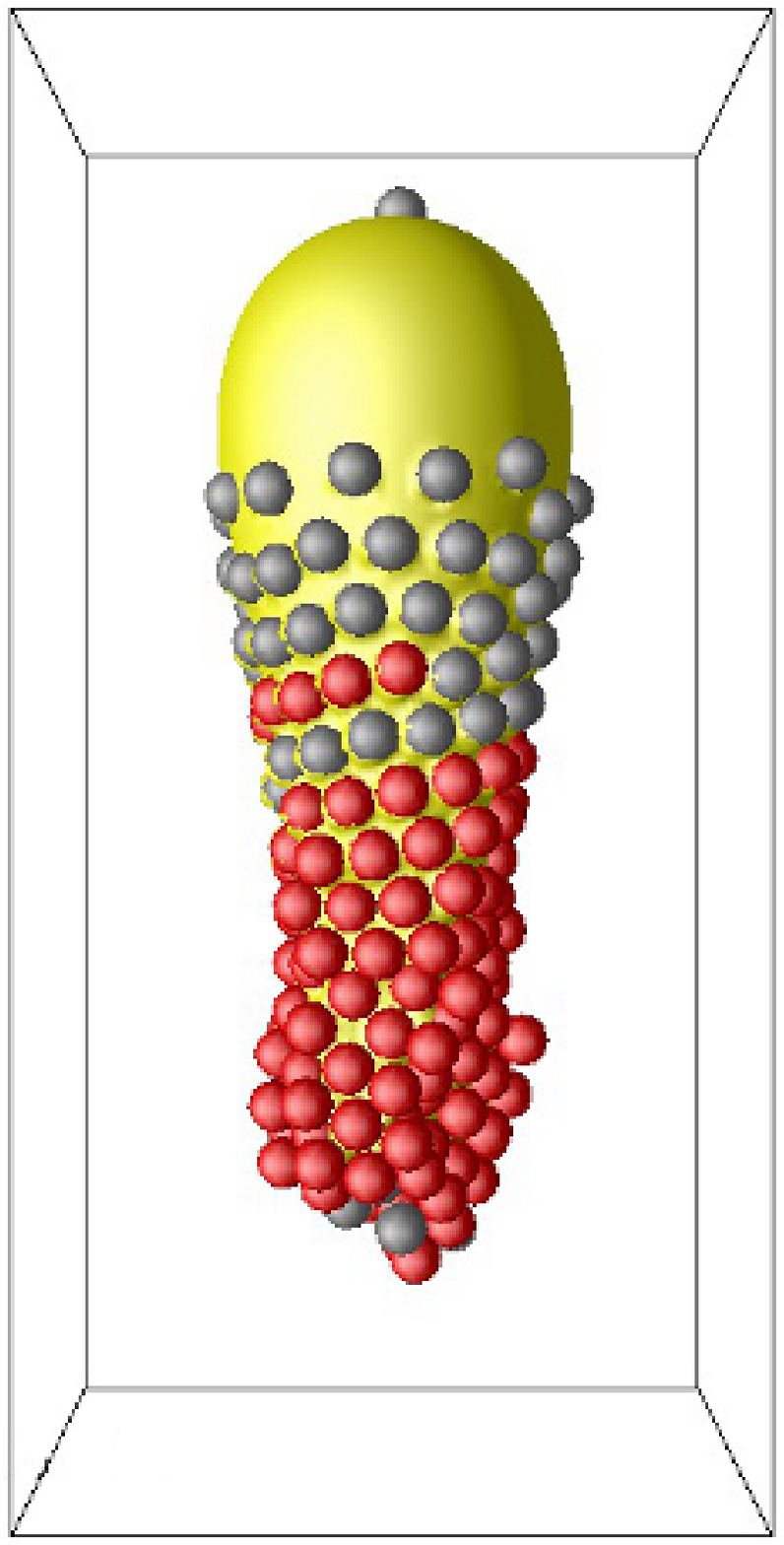}} \\
\subfigure[]{\label{fig:rN700initial}
\includegraphics[height=0.2\textwidth]{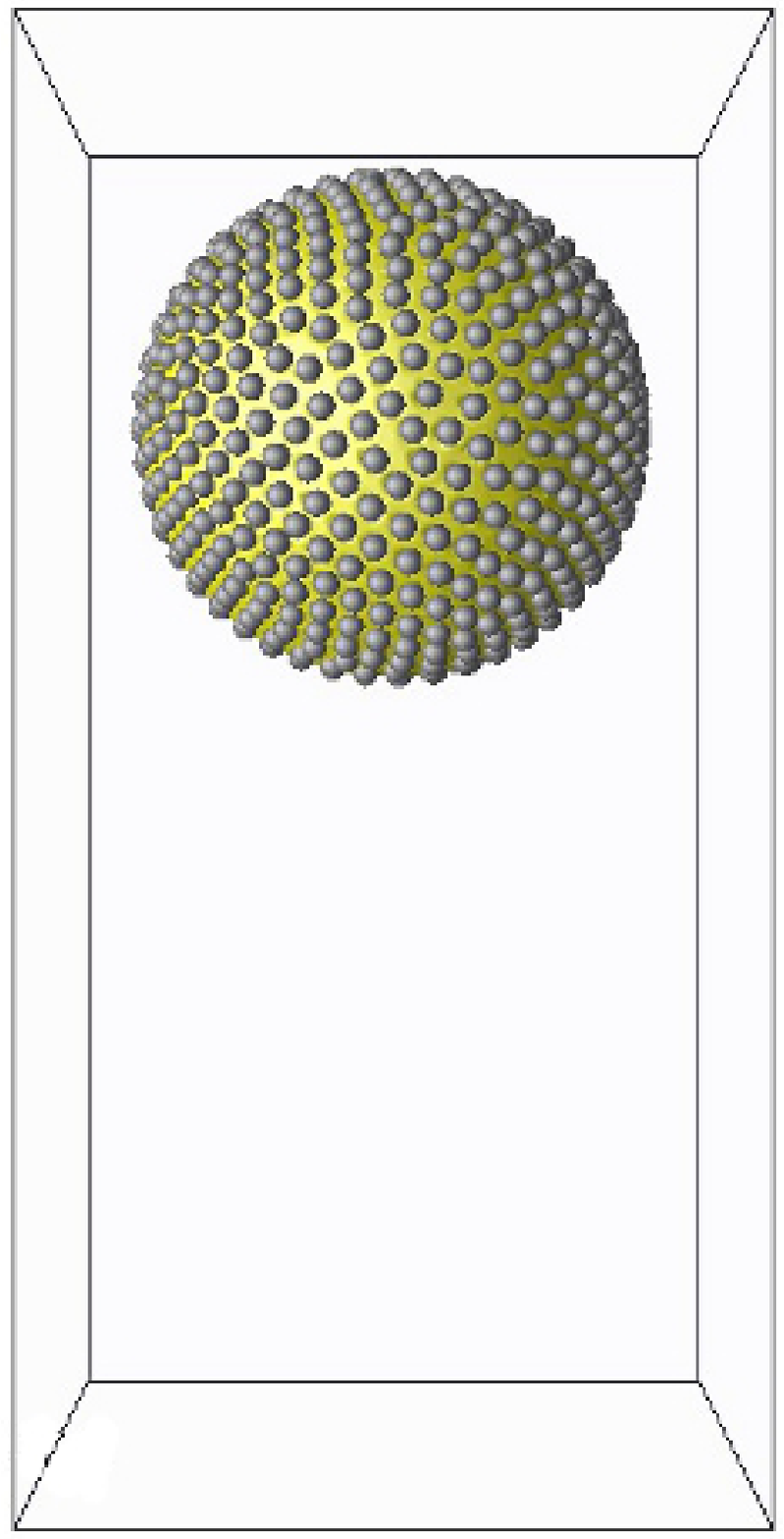}}
\subfigure[]{\label{fig:rN700_stable01}
\includegraphics[height=0.2\textwidth]{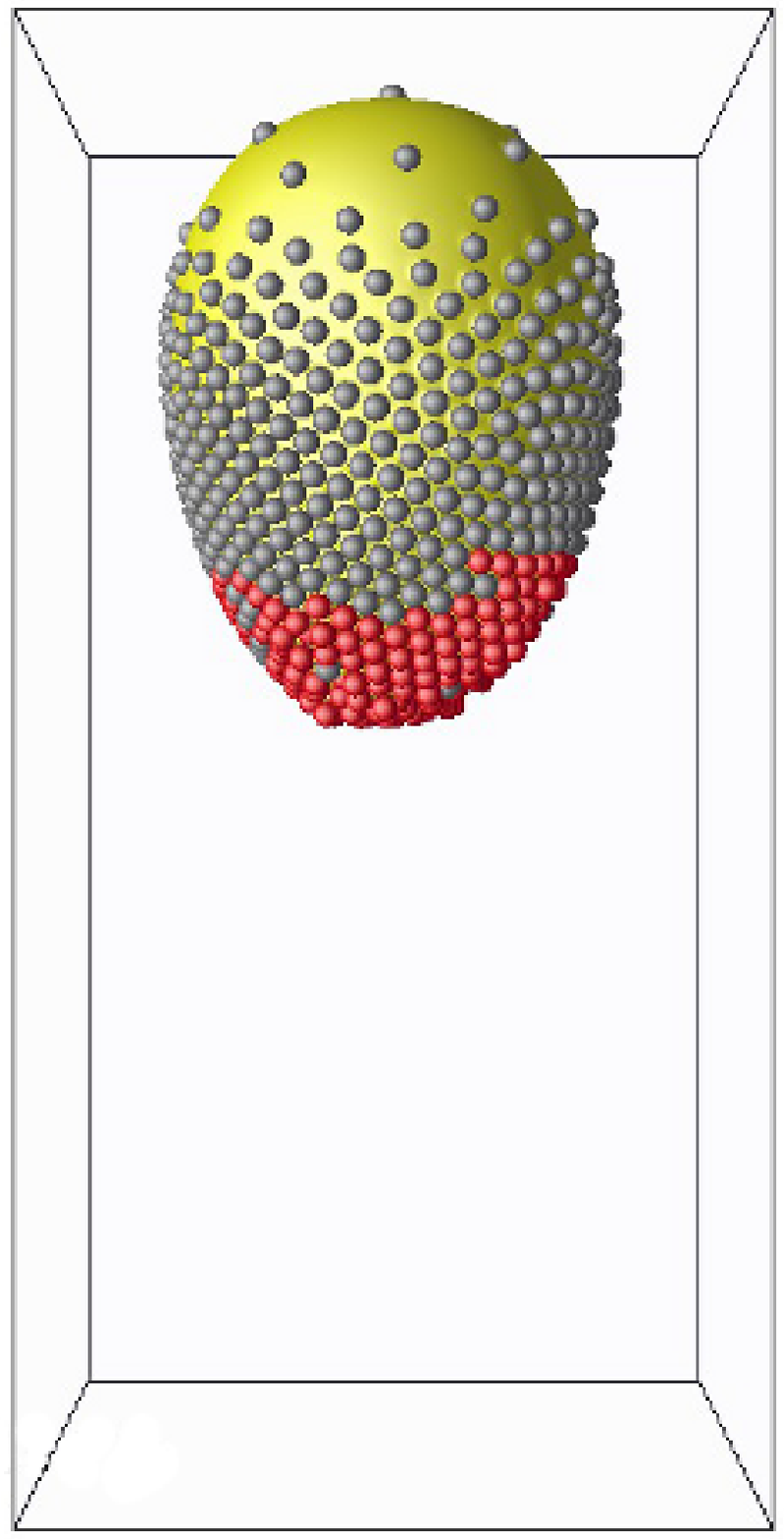}}
\subfigure[]{\label{fig:rN700_stable02}
\includegraphics[height=0.2\textwidth]{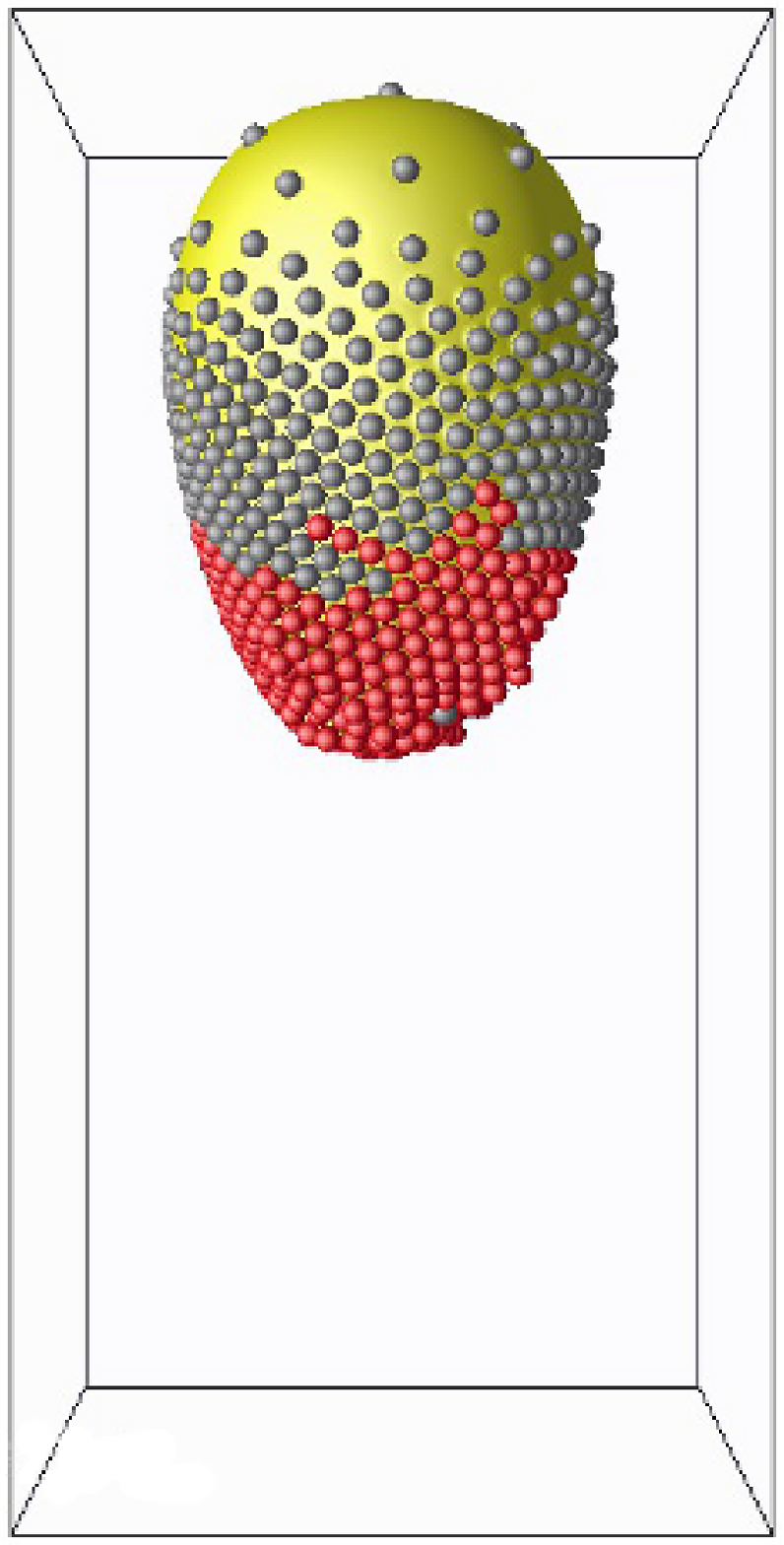}}
\subfigure[]{\label{fig:rN700_unstable}
\includegraphics[height=0.2\textwidth]{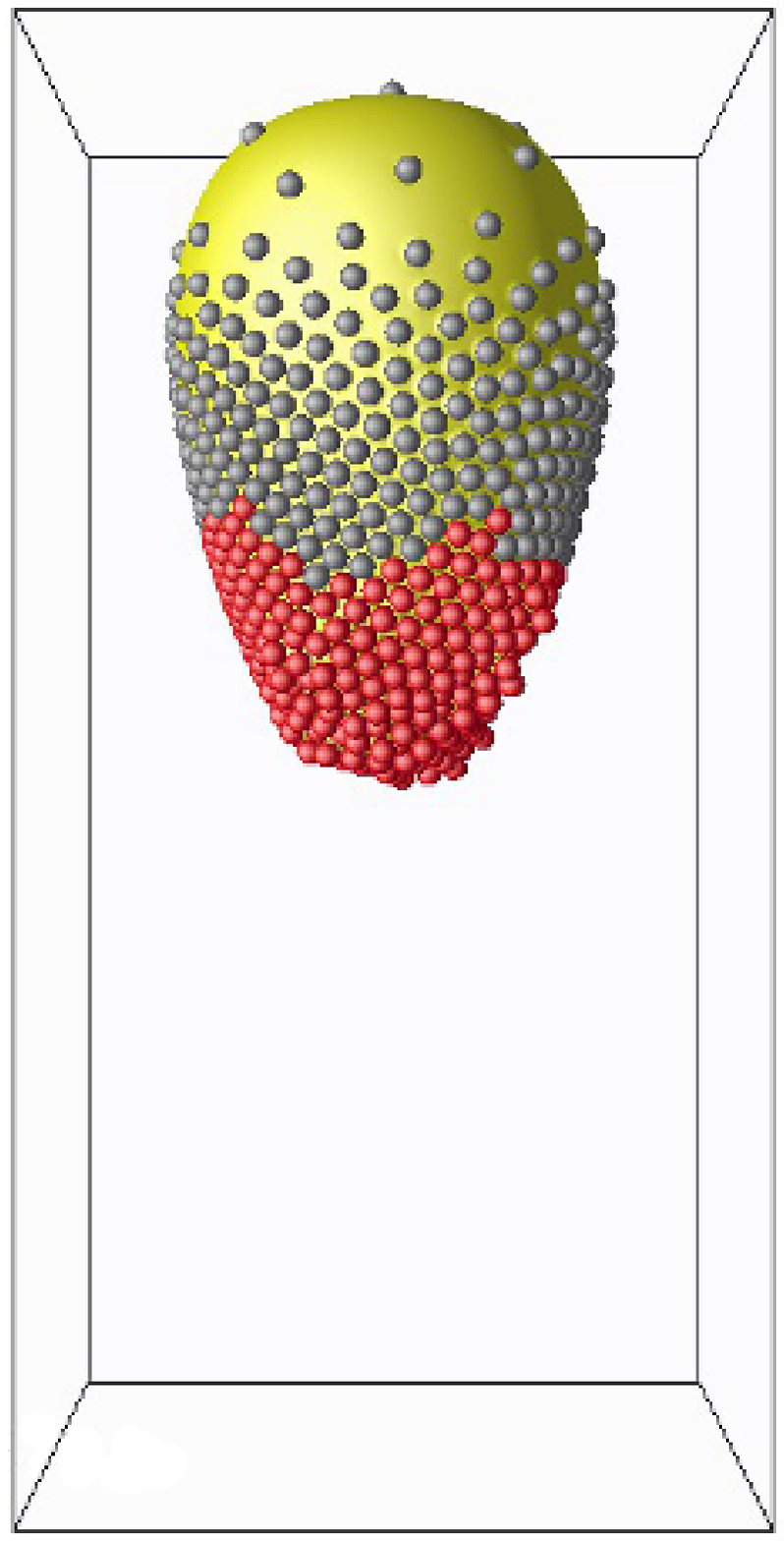}} 
\caption{(Color online.) \small{(a-h): Snapshots at $N=194$. (a) The ordered initial configuration; (b)-(d) are snapshots at $t=37t_S$ evolved from (a) with $\rm Bo=0.050, 0.051$ and $0.052$ respectively. Stable states are shown in (b) $\rm Bo=0.050$ with $N_{c}=80$, and (c) $\rm Bo=0.051$ with $N_{c}=78$. (d) Unstable state found at $\rm Bo=0.052$ with $N_{c}=121$ at $t=t_d=37t_S$.
(e) The disordered initial configuration; (f)-(h) are snapshots at $t=30t_S$ evolved from  (e) with $\rm Bo=0.050, 0.053$ and $0.055$ respectively. Stable states are shown in (f) $\rm Bo=0.050$ with $N_{c}=74$, and (g) $\rm Bo=0.053$ with $N_{c}=80$. (h) Unstable state found at $\rm Bo=0.055$ with $N_{c}=122$ at $t=t_d=30t_S$.
(i)-(l): Snapshots for N=700. (i) The disordered initial configuration; (j)-(l) are snapshots at $t=14.5t_S$ from the configuration (i) with $\rm Bo=0.030, 0.033$ and $0.034$ respectively. Stable states are shown in (j) $\rm Bo=0.030$ with $N_{c}=193$, and (k)$\rm Bo=0.033$ with $N_{c}=315$. (l) Unstable state found at $\rm Bo=0.034$ with $N_{c}=342$ at $t=t_d=14.5t_S$. }
Particles separated by $h<0.04a$ from a neighbor are shown in red.}
\label{fig:stable/unstable}
\end{figure}

\textit{Results:}
We define the critical Bond number \Bocm $(N,r)$ as the threshold beyond which one or more particles break away from a droplet covered by $N$ particles. For Bo $<$ \Bocm $(N,r)$ the droplet instead achieves a final steady state that is distorted but not ruptured. Since we work at constant $\alpha = N/4r^2 \simeq 0.53$, we now drop the argument $r$ and refer simply to \Bocm $(N)$.
In practice we find particle ejection always precedes disconnection of a droplet into two pieces and therefore used the time of first detachment, $t_d$, as a measure of when breakup occurs. (As time unit we adopt the Stokes time $t_S =6\pi\eta a^2/F$.)

Starting from an equilibrated spherical droplet and switching on the external forces, particles start to move downwards (while the buoyancy force holds the droplet up). Upper sections of the droplet become bare \cite{fuller:2005/a,Joe} and a dense interfacial `sediment' is created towards the bottom of the droplet. As the sediment builds, forces are transmitted to the particles beneath. The lower part of the droplet can then become highly distorted. 
In contrast to droplets initiated from a disordered configuration, which maintain a prolate spheroidal shape throughout the stable regime, the ordered arrangements give almost symmetric force patterns in which the most unstable particle lies at the tip of a cusp that develops at the droplet bottom. For Bo $<$ \Bocm $(N)$, such a cusped configuration can remain stable indefinitely. Fig.~\ref{fig:stable/unstable} shows snapshots for both stable and unstable 
cases 
at the time $t_d$. Shown in red are those particles whose interparticle separation $h<0.04 a$. This criterion identifies the main load-bearing particles in the system \cite{supmat}.

Although the body force changes only by two percent between the highest stable and lowest unstable value examined in Fig.~\ref{fig:stable/unstable}, there is a clear difference in their configurations at the detachment time $t_d$. In both cases, 
close-packed clusters of $N_c$ load-bearing particles are observed at the bottom of droplets; while the geometry of these regions depends on initial conditions as detailed above, in the last stable droplet configuration  $N_c\simeq 0.4N$ for the ordered and disordered cases. In the unstable state, immediately prior to particle ejection, $N_c\simeq 0.6N$ is rather larger but again nearly independent of the initial condition.
In Table \ref{table:N194unstable}
we report $N_{c}(t_d)$ for several values of Bo just above the threshold, in the case of ordered (square) and disordered packings with $N=194$.  We find $N_{c}>N/2$ in all unstable cases, with rather little dependence on Bo (once the critical value is exceeded). As expected, however, a stronger body force (larger Bo) causes the detachment time $t_d$ to decrease markedly. 

\begin{table}
\begin{center}
\begin{tabular}{c c c c }
\hline\hline
Initial state & $\rm Bo$    & $t_{d}/t_S$ & $N_{c}(t_d)$ \\ \hline
Ordered         & $0.052^\dag$ & $37$  & 121 \\
              & $0.053$ & $27$  & 125 \\
              & $0.054$ & $21$   & 114 \\ 
              & $0.055$ & $18$   & 130 \\ \hline
Disordered      & $0.055^\dag$ & $30$   & 122 \\ 
              & $0.060$ & $15$   & 126 \\
              & $0.065$ & $10$   & 127 \\ \hline
        
\end{tabular}   
\caption[smallcaption]{The time $t_{d}$ of first particle detachment and the size $N_{c}$ of the load-bearing cluster immediately prior to $t_d$ for unstable droplets of $N=194$ with ordered and disordered initial states. Superscript $^\dag$ denotes \Bocm $(N)$ (numerically, the smallest Bo found for which breakup occurs).}
\label{table:N194unstable}
\end{center}
\end{table}

\begin{figure}[tbph]
\centering
\vspace{.15in}
{\includegraphics[width=0.45\textwidth]{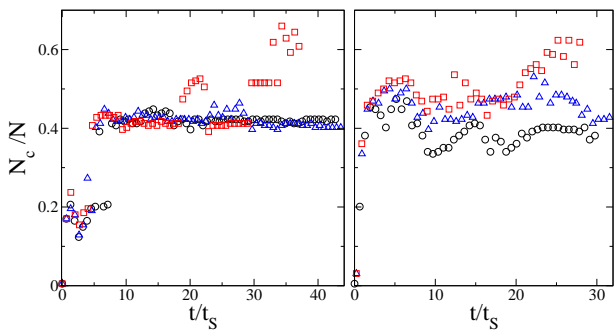}}
\caption{(Color online.) The evolving cluster size, $N_{c}(t)$ for the simulations shown in Fig.~\ref{fig:stable/unstable} (a-h).  Left: $N_{c}(t)$ for the ordered initial configuration: black circles, $\rm Bo=0.050$, blue triangles $\rm Bo=0.051$, red squares $\rm Bo=0.052$. Right:  disordered initial configuration: black circles  $\rm Bo=0.050$, blue triangles $\rm Bo=0.053$, red squares $\rm Bo=0.055$. For unstable cases, the plots end upon particle ejection.}
\label{fig:Ncluster}
\end{figure}

Fig.~\ref{fig:Ncluster} depicts the time evolution of $N_{c}$ in the cases with $N=194$. This shows the buildup of the sedimented cluster after switching on gravity, followed by a period of saturation. Beyond $t=13t_S$, stable droplets reach steady state with roughly constant $N_{c}$, although disordered cases show fluctuations due to reorganization of the randomly packed 2D sediment. In the unstable cases, such reorganization is always seen during the plateau region; following a series of these plastic events $N_{c}$ starts increasing steadily until breakup occurs.

\begin{figure}[tbph]
\centering
\vspace{.1in}
\includegraphics[width=0.3\textwidth]{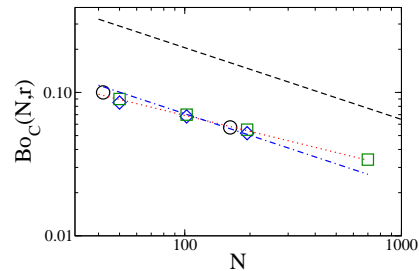}
\caption{(Color online.) Log-log plot of critical Bond number for a droplet covered with $N$ particles, \Bocm $(N)$, vs $N$. Blue diamonds: square ordered; green squares: disordered; black circles: hexagonal ordered. Blue dot-dashed line: best fit of all data to  ${\rm Bo}_{\hbox{\rm c}}(N) = A N^{-1/2}$ ($A = 0.71$). Red dotted line: best fit to ${\rm Bo}_{\hbox{\rm c}}(N) = A N^{B}$ ($A = 0.38; B=-0.37$). Black dashed line: bound found from $\lambda=3/2$ (see text).}
\label{fig:scalingN0.5}
\end{figure}

\textit{Scaling of critical Bond number:}
The above mechanism entails asssembly of a close-packed loadbearing cluster, involving roughly half the particles, which either maintains itself indefinitely (Bo $<$ \Bocm $(N)$) or, after a series of plastic rearrangements, 
finally gives way, causing particle ejection and droplet rupture (Bo $>$ \Bocm $(N)$). The droplet prior to this event is distorted significantly; its shape, while dependent in detail on the initial condition, is roughly independent of $N$, and involves order-one departures from a sphere. There is no sign of extreme cylindrical elongation as can arise in shear flow \cite{Bentley,Lyn}.

Although breakup is preceded by particle ejection, with all parameters studied here we do not see a stream of particles being ejected from the lower tip of the droplet as might be expected were the sediment of loadbearing particles to behave as a fluid. Such an outcome would allow the weight of all $N$ particles to be borne by a single `keystone' particle at the bottom of the droplet. Ejection of that particle, possibly nucleating an instability of the entire structure, could then be expected at a critical Bond number scaling as \Bocm $\sim$\Bocm $(1,r)/N$ \cite{kim:one}. For our nearly hard-sphere particles\cite{supmat}, this is not observed here \cite{Foot2}; we instead find a scaling closer to $N^{-1/2}$ (see Fig.~\ref{fig:scalingN0.5}). This is because the cluster has formed a jammed structure in which forces are transmitted laterally as well as vertically. This semi-rigid assembly then becomes collectively unstable (see Movies at \cite{supmat}).

Given the destabilization mode observed, we now estimate \Boc by addressing the force balance on the close-packed sediment, viewed as a solid cluster. (See \cite{Oettel,Stone} for related arguments.) This has weight $FN_c\simeq FN/2$ and to a first approximation occupies the lower half of a droplet that is only moderately distorted. Accordingly at Bo $=$ \Boc this weight is just balanced by the surface tension force acting vertically across the equator of the deformed droplet, $2\pi\sigma \tilde R$ where $\tilde R \simeq R/2$ is the radius at the equator. Bearing in mind that the initial surface coverage (here held constant) obeys $\alpha = Na^2/4R^2 = N/4r^2$ we then have the estimate \Bocm $(N,r) \simeq 3\tilde R/Na$ or, introducing a negotiable geometrical constant $\lambda \simeq 0.75$,
$
{\rm Bo}_{\hbox{\rm c}}(N,r)  = {\lambda}/{\sqrt{\alpha N}}.
$
The predicted scaling \Bocm $\sim N^{-1/2}$ is tested in Fig.~\ref{fig:scalingN0.5} which plots all data for \Boc against $N$ on a log-log plot. The best fit (treating data for all initial conditions equally)  has $\lambda = 0.52$. 

Table \ref{table:criticalBo} shows all the \Boc values found in this study alongside run-by-run values of $\lambda$. The residual increasing trend of $\lambda$ with $N$ is consistent with the fact that the pre-breakup droplet shape becomes more spherical for larger $N$. Indeed the quantity $\lambda R/\tilde R$, with $\pi\tilde R^2$ defined as the cross sectional area of the deformed droplet at the height of the uppermost load-bearing particle, has no residual trend (see Table \ref{table:criticalBo})  \cite{Footnote}. A free fit to the exponent gives \Bocm $\sim N^{-0.37}$, or $\tilde R/R\sim N^{0.13}$. We have no explanation for this weak residual trend, but note that it cannot persist to very large $N$ unless droplets become oblate. Excluding that outcome, a bound \Bocm $\ge 3/(2\sqrt{\alpha N})$ at large $N$ is found by balancing the tensile force $2\pi\sigma R$ in an undeformed spherical configuration against the weight $FN/2$ of its lower hemisphere. (Any prolate deformation reduces the tensile force, decreasing \Boc.) This bound is plotted in Fig.~\ref{fig:scalingN0.5}; extrapolating our $N^{-0.37}$ fit, saturation is expected only at $N\simeq 10^4$. 
 
\begin{table}
\begin{center}
\begin{tabular}{c  r c r c c r r r r}
\hline
 & $N$ & ${\rm Bo}_{\hbox{\rm c}}(1,r)$ & ${\rm Bo}_{\hbox{\rm c}}(N)$ & $N_{c}(t_d)$ & $t_{d}/t_S$ & $\lambda$ & $\tilde R/R$ & $\lambda R/\tilde R$\\ \hline
S & $50$ & $0.55$ & $0.085$ & $29$ & $103$ & $0.44$  & $0.52$ & 0.84\\
        & $102$& $0.57$ & $0.068$ & $54$ & $38.5$ & $0.50$  & $0.65$ & 0.77\\
        & $194$& $0.60$ & $0.052$ & $121$ & $37$& $0.53$  & $0.55$ & 0.97\\ \hline
H & $42$ & $0.54$ & $0.10$ & $23$ & $38$ & $0.47$  & $0.49$ & 0.95\\
& $162$& $0.59$ & $0.057$ & $88$ & $38.5$ & $0.53$  & $0.69$ & 0.77\\ \hline
D& $50$ & $0.55$ & $0.090$ & $31$ & $59$ & $0.46$ &  $0.63$ & 0.73\\
        & $102$& $0.57$ & $0.070$ & $52$ & $21$ &  $0.52$  & $0.59$ & 0.88\\
        & $194$& $0.60$ & $0.055$ & $122$ & $30$& $0.56$  & $0.60$ & 0.94\\ 
        & $700$& $0.63$ & $0.034$ & $342$ & $14.5$& $0.65$  & $0.77$ & 0.85\\ \hline
\end{tabular}
\caption[smallcaption]{Critical Bond numbers (\Bocm $(N)$ for a droplet coated with $N$ particles at initial coverage $\alpha \simeq 0.53$, and \Bocm $(1,r)$ for a single particle on a droplet of the same radius); size of the load-bearing cluster prior to breakup; detachment time; breakup parameter $\lambda$; and shape parameter $\tilde R/R$. Initial configurations S, square, H, hexagonal, D, disordered. }
\label{table:criticalBo}
\end{center}
\end{table}

\textit{Conclusion:}
We have investigated by LB simulations the destabilization of an emulsion droplet coated with hard spheres under a body force that acts equally on all the particles
\cite{fuller:2005/a,aubry:2010/a,Joe}. We confirmed that the force required for destabilization is far less than would be expected if the force were to cause independent detachment of the particles from the fluid-fluid interface \cite{kim:one}. However, it is also far greater than would be expected if the weight of all $N$ particles were to accumulate as a net detachment force on a single `keystone' particle at the bottom of the droplet. That picture, which implicitly considers the stabilizing particle layer to comprise a 2D fluid, must now be replaced by one in which the lower part of that layer has solidified. A simple geometrical analysis then suggests a critical Bond number scaling as $N^{-1/2}$ which gives a satisfactory first account of the simulation data.

\textit{Acknowledgments:}
EGK thanks Siegfried Dietrich for hospitality in Stuttgart. We thank Jan Guzowski and Joe Tavacoli for useful discussions. MEC is funded by the Royal Society.

\newpage

\section{Field-Induced Breakup of Emulsion Droplets Stabilized by Colloidal
Particles: Additional Online Material}

\section{Simulation method} \label{method}

All simulations are performed using the LB method on a $D3Q19$ lattice for
a binary fluid, with the bounce-back-on links method used to handle momentum exchange with solid particles \cite{Bounce}, as extended to the binary fluid case in \cite{stratford:2005/a}. The fluid-fluid interfacial tension and width are determined by the parameters of Ginzburg-Landau free energy functional $F[\psi]= \int dV \left( \frac{A}{2}\psi^2 + \frac{B}{4}\psi^4 +  \frac{\kappa}{2}(\nabla \psi)^2 \right)$, where $\psi$ is a compositional order parameter.
Following \cite{kendon:2001/a,kim:one} we express these parameters in lattice units (LU) as $-A=B=0.0011832$ and $\kappa=0.0023664$; the interfacial thickness and the nominal interfacial tension are deduced as $\xi=2.0$ LU and $\sigma_0=1.5776\times10^{-3}$ LU respectively. The fluid viscosity is set at $\eta=0.1$ LU. Lattice units are as defined in \cite{kim:one} but for our purposes are arbitrary: all that matters is the Bond number. Note that the Reynolds number, which controls the relative importance of inertia compared to viscosity, remains small throughout our simulations which represent the Stokes flow limit \cite{kim:one}. The sedimentation velocity of one particle $v_s$ can be used to define a capillary number Ca $=\eta v_s/\sigma = 2{\rm Bo}/9$ which remains small in all our simulations.
 
From LB benchmark tests with this parameter set, the numerical value of surface tension is determined as $\sigma = 1.51 \times 10^{-3}$ LU which is 4\% less than the nominal value $\sigma_0$. (Discrepancies of this scale are common in LB \cite{kendon:2001/a}.) Monodisperse neutral-wetting particles are introduced with hydrodynamic radius $a_h=3.9$ LU. This choice is imposed by the numerical resources available; note however that $a_h$ is only twice the interfacial width $\xi$. For that reason, one expects some discrepancy between the values of \Bocm $(1,r)$ seen in simulations and those arising for a structureless interface ($\xi\to 0$). The values tabulated in the main text are those found by numerical simulations on a single particle residing on a droplet of radius $R = ra$. We have however confirmed that for a flat interface ($r\to\infty$), our methods recover the expected result: indeed we find \Bocm $(1) = 0.74\simeq 3/4$. Particle ejection is defined by the surface-average value of $\psi$ becoming sufficiently different from zero.

Interparticle forces are mediated by a soft-core potential $U^{sc}(h)$ chosen to prevent the surface-to-surface interparticle separation $h$ from becoming too small. (To model capillary forces correctly, it is important that the lattice Boltzmann nodes surrounding each particle mainly represent fluid not solid sites.) We choose $U^{sc}(h)=u(h)-u(h_c)-(h-h_c)(\frac{du}{dh})_{h=h_c}$, where $u(h)=\gamma (h_0/h)$, and $h_c$ is the maximum range of the interaction; we select $\gamma=10^{-4}$, $h_0=0.1$, and $h_c=0.25$ \cite{kim:one}. In practice, the particles behave as nearly impenetrable spheres of radius $a \simeq a_h\simeq 4$ LU. Finally, to allow for the discretization error in the hydrodynamic forces at small separations, a lubrication force (proportional to the normal component of the interparticle velocity) acts whenever $h<0.3$ LU. The criterion $h < 0.15$ LU $= 0.04a$ is used to identify load-bearing particle contacts; this lies within the range of the soft-core potential defined above.

\section{Additional Results}

\begin{figure}[tbph]
\centering
\vspace{.5in}
\subfigure[]{\includegraphics[height=0.3\textwidth]{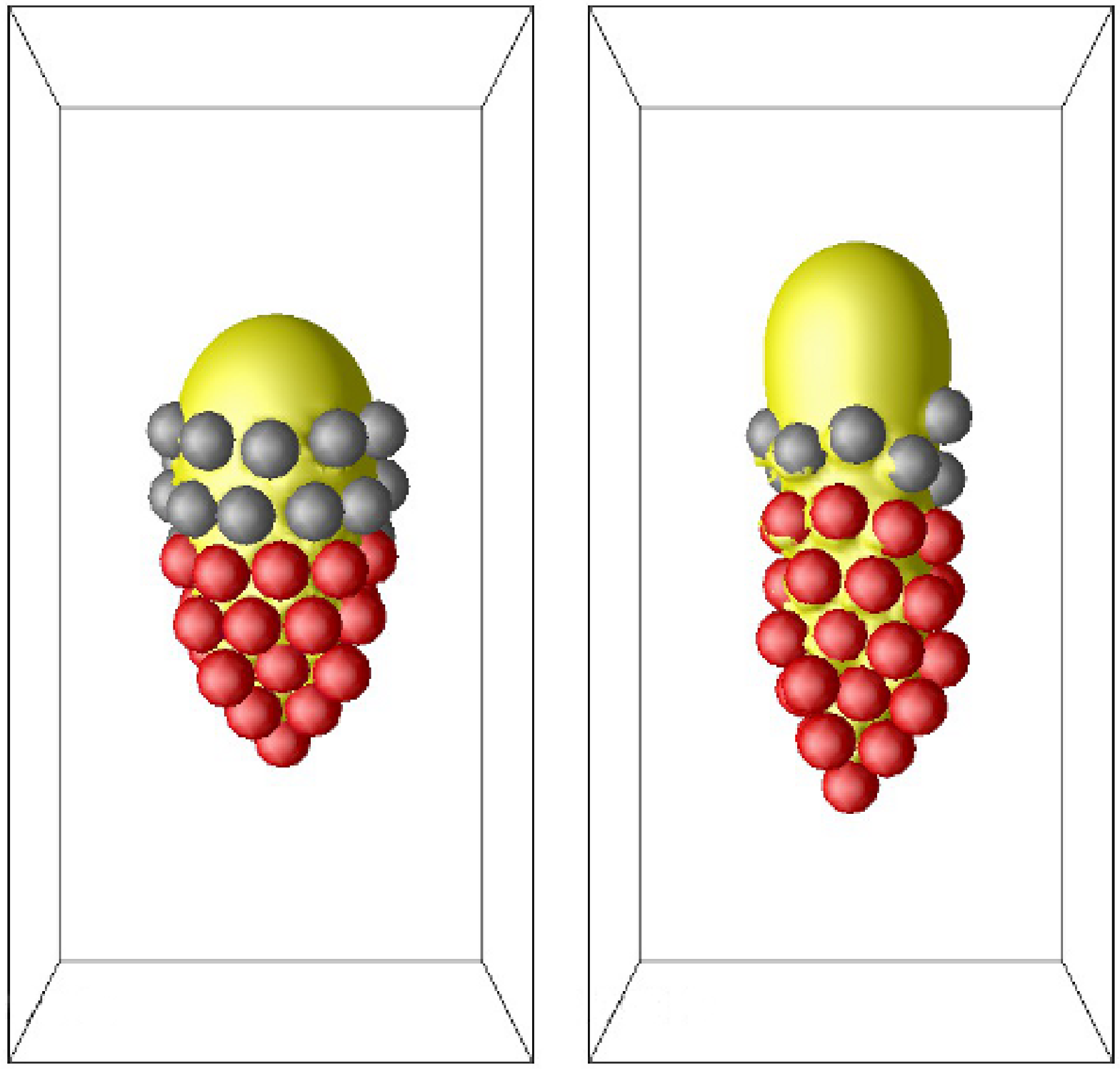}}
\subfigure[]{\includegraphics[height=0.3\textwidth]{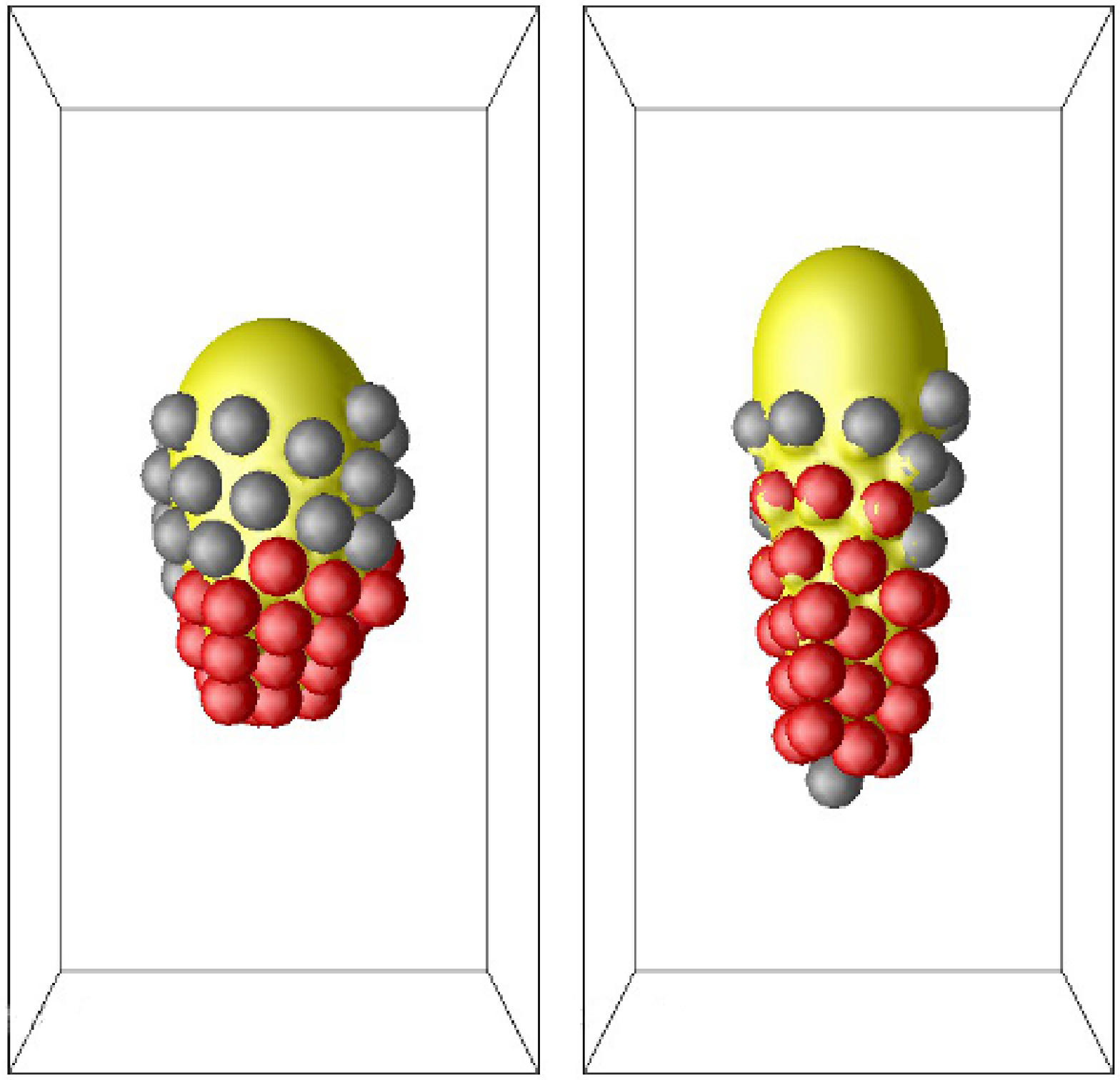}}
\vspace{.05in}
\caption{Comparison of particle-packing between ordered and disordered initial configurations at different time steps. (a) Snapshots of a droplet at $t=89 t_S$ on the left and at $t_{d}=100 t_S$ on the right from initial ordered configuration. (b) Snapshots of a droplet at $t=51 t_S$ (left) and at $t_{d}=59 t_S$ (right) from disordered configuration.}
\label{fig:order/random}
\end{figure}

Figure \ref{fig:order/random} shows snapshots of the particle configurations in much smaller droplets ($N=50$), prepared initially in ordered (square lattice) and disordered states, at two different later times. These are for the lowest unstable Bo ($\simeq$ \Bocm $(N)$) so that both droplets will subsequently break. The left pictures in Figures \ref{fig:order/random}(a) and \ref{fig:order/random}(b) are configurations where $N_{c}$ has reached the near-saturation plateau value (compare Fig.~\ref{fig:Ncluster}) showing semi-crystalline clusters at the droplet bottom. The right pictures in Figures \ref{fig:order/random}(a) and \ref{fig:order/random}(b) are the configurations immediately before the first particle detachment. In each case there is a neck close to the droplet equator, with load-bearing particles below the neck. Although $N$ here is quite small, these configurations are remarkably similar to those seen in Fig.~\ref{fig:stable/unstable} for $N=194$ and even $N = 700$. 

\begin{figure}[tbph]
\centering
\subfigure[]{\label{fig:aH7.8_initial}\includegraphics[height=0.2\textwidth]{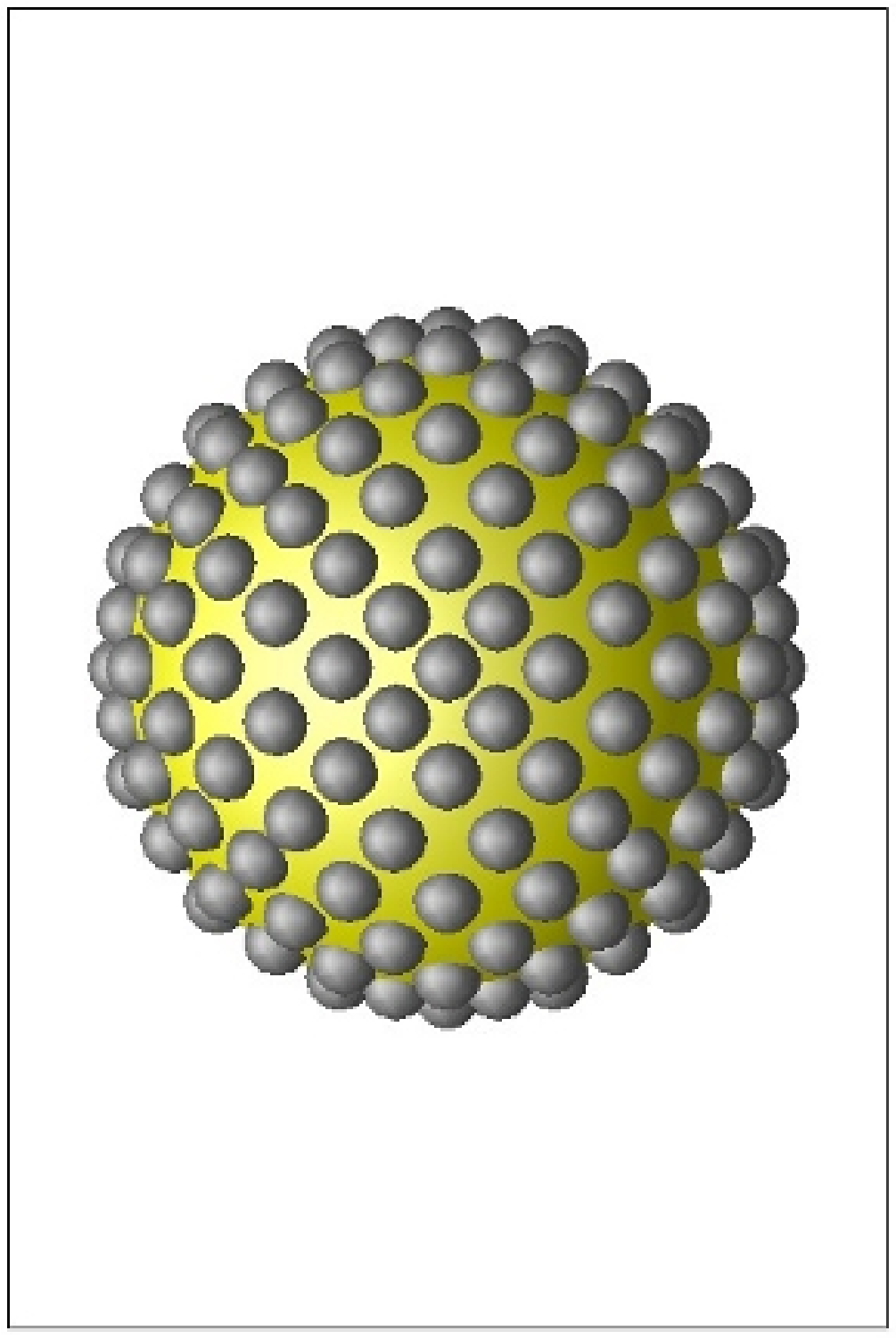}}
\subfigure[]{\label{fig:order_stable01A}\includegraphics[height=0.2\textwidth]{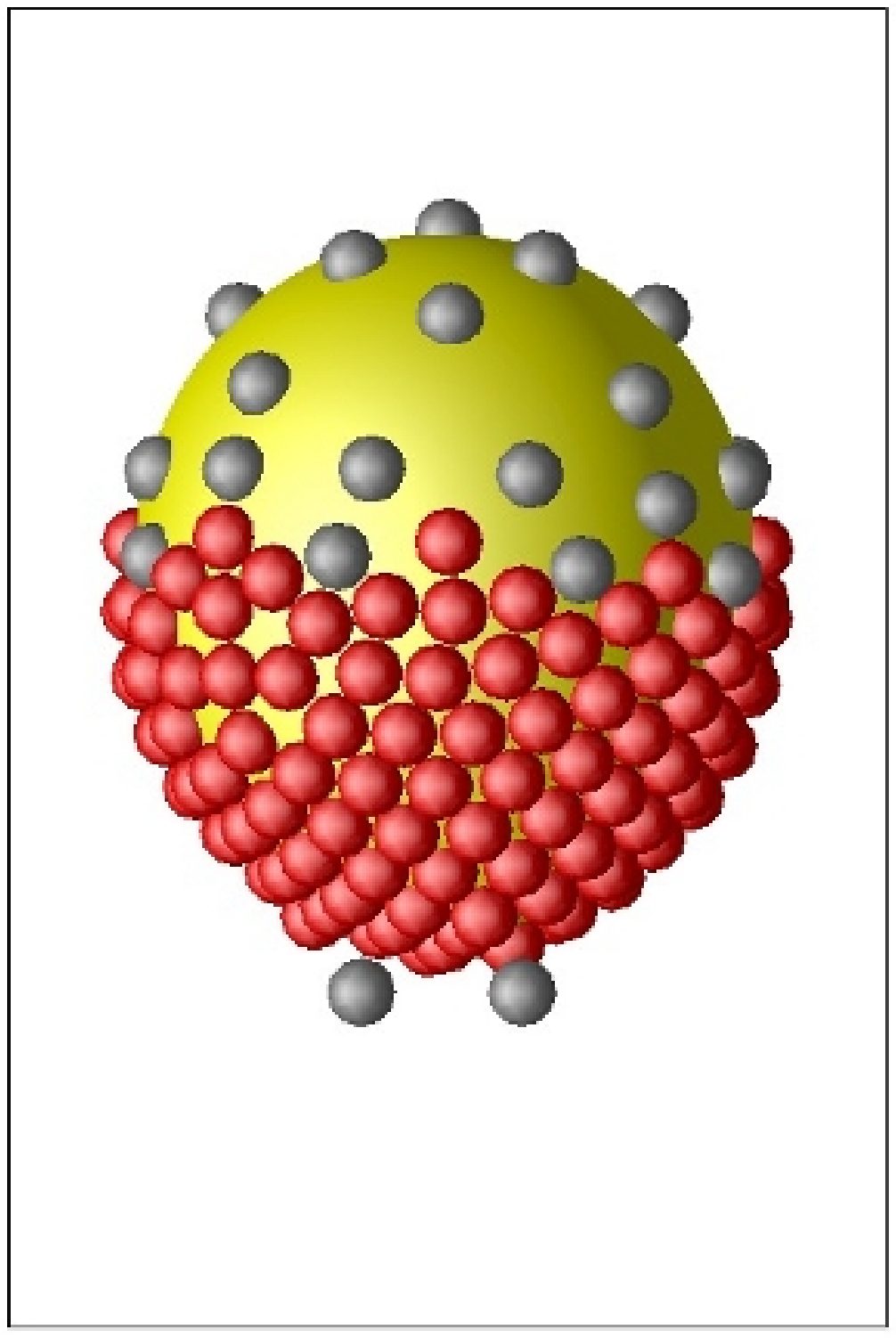}}
\subfigure[]{\label{fig:order_stable02A}\includegraphics[height=0.2\textwidth]{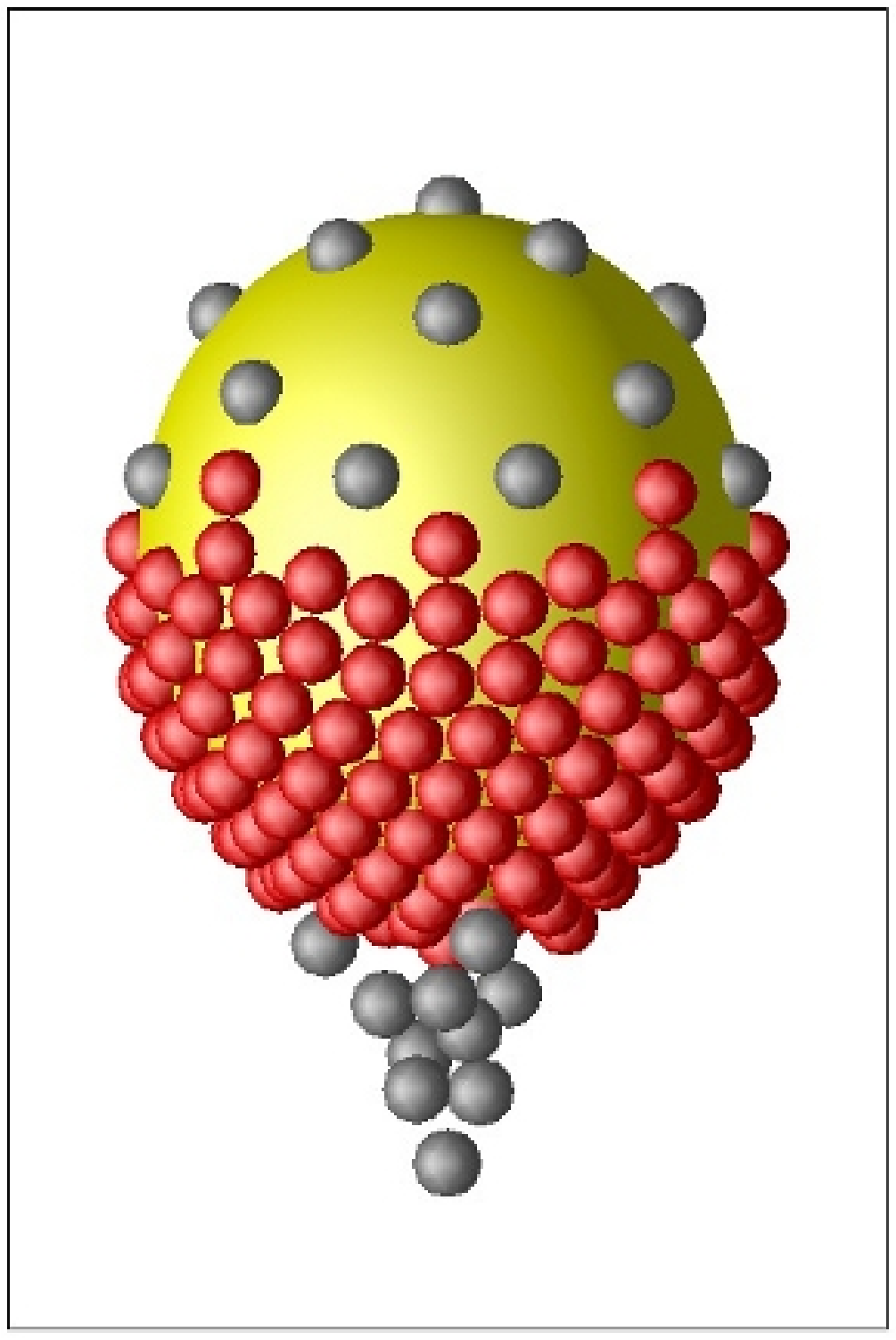}}
\subfigure[]{\label{fig:order_unstableA}\includegraphics[height=0.2\textwidth]{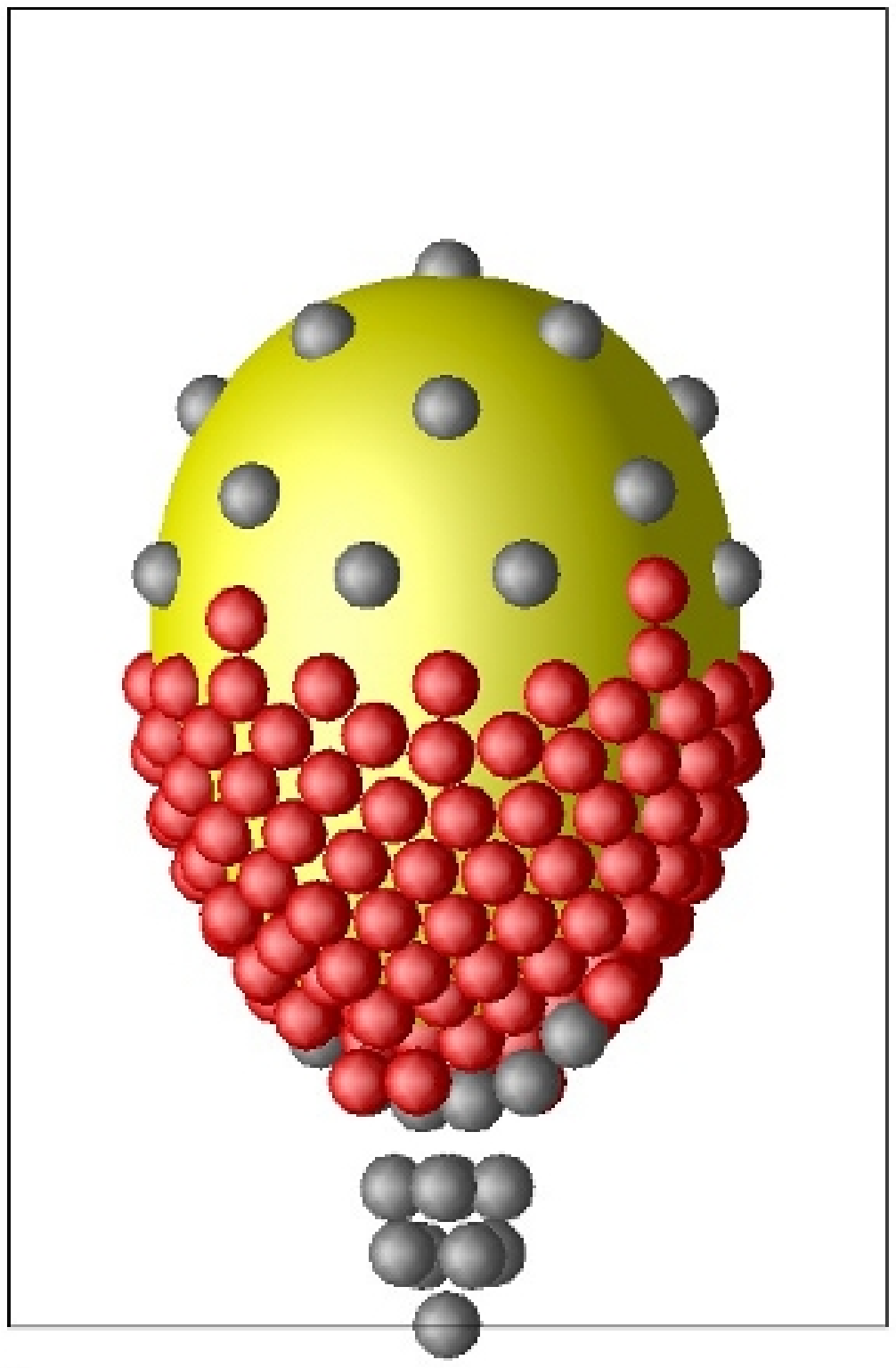}}\\
\caption{Snapshots for droplets at $N=194$ with hard core interaction radius twice the capillary radius, showing keystone detachment mode in place of the collective breakup. (The capillary radius is used in the definition of the Bond number; the hard core radius is used for the visualization.) (a) The ordered initial configuration; (b)-(d) are snapshots at $t=11.4t_S$ evolved from (a) with $\rm Bo=0.048, 0.064$ and $0.11$ respectively. (b) $\rm Bo=0.048$ with $N_{c}=157$ and $N_{f}=2$, (c) $\rm Bo=0.064$ with $N_{c}=160$ and  $N_{f}=8$. (d) $\rm Bo=0.11$ with $N_{c}=157$ and $N_{f}=21$.}
\label{fig:N194aH7.8/unstable}
\end{figure}

Figure \ref{fig:N194aH7.8/unstable} shows the breakup mode for an emulsion droplet ($N=194$) in which the hard-core interaction radius of the particles is unchanged but the radius they occupy on the interface is halved. For a given gravitational force per particle, this does not alter the force balance argument for the collective detachment of the lower hemisphere. However it does reduce four-fold the maximum total force that can be transmitted to the bottom-most particle without this detaching individually from the interface. In this case, a `keystone' type failure mode is seen, in which particles stream off the bottom of the droplet without entrainment of the enclosed fluid. Note that similar physics could arise for partially wetting particles with a contact angle far from 90 degrees: in this case also, the radius of the interfacial disc covered by each particle is significantly less than its hard sphere interaction radius. This again reduces the collective force sustainable without individual detachment, and therefore lowers stability with respect to the keystone detachment mode without having the same effect on the collective mode (in which a naive force balance again equates the total mass in the lower hemisphere to $2\pi R\sigma$, independent of contact angle).

\section{Movies}

The following online movies are available showing the time evolution of stable and unstable droplets for various sizes and initial conditions.

Movie 1. $N = 194$, ordered initial condition, stable (parameters as in Fig.1b).

Movie 2. $N = 194$, ordered initial condition, unstable (parameters as in Fig.1d).

Movie 3. $N = 194$, disordered initial condition, stable (parameters as in Fig.1f).

Movie 4. $N = 194$, disordered initial condition, unstable (parameters as in Fig.1h).

Movie 5. $N = 700$, disordered initial condition, stable (parameters as in Fig.1j).

Movie 6. $N = 700$, disordered initial condition, unstable (parameters as in Fig.1l).

Movie 7. $N=194$, ordered initial condition, unstable to `keystone' type failure mode (parameters in Fig. 5c).  
\vfill\eject

\begin{thebibliography}{100}
\bibitem{supmat} The simulation methodology, additional results, and movies are available online at XXXXXXXX.
\bibitem{Foot2} For hard spheres the collective breakup mode is more unstable than the keystone detachment mode. Remarkably, this can be reversed by halving the capillary radius of the particles, while maintaining the range of the hard-core interparticle forces. This barely affects force balance for the collective failure mode, but reduces four-fold the maximum force the keystone can sustain; ejection of individual particles now happens first \cite{supmat}.
\bibitem{Footnote} The deviation might also be linked to \Bocm $(1,r)$, which increases weakly with $N$; however, our collective mechanism presents no obvious role for this quantity.
\end{thebibliography}
\end{document}